\documentclass[superscriptaddress,twocolumn,showkeys,aps,prb,reprint]{revtex4-2}
 
\usepackage{graphicx}
\usepackage{dcolumn}
\usepackage{bm}
\usepackage{float}
\usepackage{color}
\usepackage{amsmath}
\usepackage{calligra}
\usepackage[T1]{fontenc}

\usepackage{soul,xcolor}
\usepackage{latexsym}
\usepackage{amsthm}
\usepackage{bbm}
\usepackage{amsfonts}
\usepackage{amssymb}
\usepackage{hyperref}
\hypersetup{colorlinks=true, citecolor=blue}
\setstcolor{red}

\newcommand{\ddelta}{\boldsymbol{\delta}}

\newcommand{\kk}{\mathbf{k}}

\newcommand{\vs}{\mathbf{s}}
\newcommand{\KK}{\mathbf{K}}
\newcommand{\GG}{\mathbf{G}}
\newcommand{\RR}{\mathbf{R}}
\newcommand{\Hcal}{\mathcal{H}}
\newcommand{\Ecal}{\mathcal{E}}
\newcommand{\Rcal}{\mathcal{R}}

\newcommand{\ssigma}{\boldsymbol{\sigma}}
\newcommand{\ttau}{\boldsymbol{\tau}}
\newcommand{\ppi}{\boldsymbol{\pi}}

\begin{document}

\title{Spin-valley locking in Kekul\'e-distorted graphene with Dirac-Rashba interactions}

\author{David A.\ Ruiz-Tijerina}
\email{d.ruiz-tijerina@fisica.unam.mx}
\affiliation{Departamento de F\'isica Qu\'imica, Instituto de F\'isica, Universidad Nacional Aut\'onoma de M\'exico, Ciudad de M\'exico, C.P. 04510, M\'exico}
\author{Jes\'us Arturo S\'anchez-S\'anchez}
\affiliation{Departamento de F\'isica, Centro de Nanociencias y Nanotecnolog\'ia, Universidad Nacional Aut\'onoma de M\'exico, Apdo. Postal 14, 22800 Ensenada, Baja California, M\'exico}
\affiliation{Instituto de Ciencias F\'isicas, Universidad Nacional Aut\'onoma de M\'exico, Cuernavaca, Morelos, 62210, M\'exico}
\author{Ramon Carrillo-Bastos}
\affiliation{Facultad de Ciencias, Universidad Aut\'{o}noma de Baja California, Apdo. Postal 1880, 22800 Ensenada, Baja California, M\'{e}xico.}
\author{Santiago Galv\'an y Garc\'ia}
\affiliation{Departamento de F\'isica, Centro de Nanociencias y Nanotecnolog\'ia, Universidad Nacional Aut\'onoma de M\'exico, Apdo. Postal 14, 22800 Ensenada, Baja California, M\'exico}
\affiliation{Instituto de Ciencias F\'isicas, Universidad Nacional Aut\'onoma de M\'exico, Cuernavaca, Morelos, 62210, M\'exico}
\author{Francisco Mireles}
\email{fmireles@ens.cnyn.unam.mx}
\affiliation{Departamento de F\'isica, Centro de Nanociencias y Nanotecnolog\'ia, Universidad Nacional Aut\'onoma de M\'exico, Apdo. Postal 14, 22800 Ensenada, Baja California, M\'exico}

\date{\today}

\begin{abstract}
 
The joint effects of Kekulé lattice distortions and Rashba-type spin-orbit coupling on the electronic properties of graphene are explored. We modeled the position dependence of the Rashba energy term in a manner that allows its seamless integration into the scheme introduced by Gamayun \emph{et al.}[\onlinecite{Gamayun_2018}] to describe graphene with Kekulé lattice distortion. Particularly for the Kekul\'e-Y texture, the effective low energy Dirac Hamiltonian contains a new spin-valley locking term, in addition to the well-known Rashba-induced momentum-pseudospin and spin-pseudospin couplings, and the Kekul\'e-induced momentum-valley coupling term. We report on the low-energy band structure and Landau level spectra of Rashba-spin-orbit-coupled Kek-Y graphene, and propose an experimental scheme to discern between the presence of Rashba spin-orbit coupling, Kek-Y lattice distortion, and both, based on doping-dependent magnetotransport measurements.

\end{abstract}

\maketitle

\section{Introduction}

In recent years,  graphene-metal hybrid systems have attracted much attention because they showcase new and exciting electronic and magnetic phenomena not present in pristine graphene \cite{grapheneonmetal}. Through proximity effects, these hybrid systems enable the modification and control of the electronic properties of pristine graphene, such as opening a gap between the valence and conduction bands\cite{graphenegap}; distorting the linear behavior of carriers at low energies\cite{graphenelinear}; and introducing the anomalous, spin and spin-quantum Hall effects\cite{AHE,SHE,QSHE}. Researchers aim to break graphene's symmetries---from which many of its extraordinary properties originate---through various mechanisms, to turn graphene into a suitable candidate for spintronic applications \cite{spintronicapp}. Two of the most actively researched mechanisms for breaking graphene's symmetries are lattice deformations, and spin-orbit effects induced by proximity\cite{graphenespintronics,lin2017competing,curvedgraphne,kekstrain,alex2019}. Throughout this paper, we shall focus on these two mechanisms, combining specifically Kekulé lattice distortions, and proximity-induced Rashba spin-orbit (RSO) coupling.

On the one hand, a Kekulé Y-shaped bond pattern (Kek-Y) was experimentally obtained in 2016 by Gutierrez \emph{et al.}\cite{GutierrezNature,Gutierrezdoc} by growing graphene epitaxially on a Cu(111) surface, and attributed to the commensurate lattice constants of graphene and the copper substrate, combined with the presence of copper vacancies acting as ``ghost'' adatoms. Eom and Koo\cite{eom2020} observed both Kek-Y and Kek-O textures by inducing nanoscale strain on graphene using a silicon dioxide substrate. Finally, the graphene Kek-O texture has been observed by Li \emph{et al.} in quantum Hall ferromagnetic states\cite{PhysRevB.100.085437-Hall}, and more recently by Bao \emph{et al.} in Li-intercalated graphene samples\cite{PRL.126.206804-BAO2021,PRB.105.L161106-BAO2022}, and by Qu \emph{et al.} in graphene decorated with Li adatoms\cite{Science2022-Ubiquitous}. In order to determine the electronic structure of graphene with Kekulé lattice distortion, Gamayun \emph{et al.}\cite{Gamayun_2018} derived low-energy Dirac Hamiltonians for both the Kek-Y and Kek-O textures. In the former case, they found that the lattice distortion introduces a new coupling between the electronic momentum and its valley isospin, as a result of the broken chiral symmetry. The low-energy spectrum preserved its linear behavior near the Fermi level, but with the valley degeneracy breaking resulting in two Dirac bands with distinct Fermi velocities.

On the other hand, a Rashba-type spin-orbit coupling has been confirmed in graphene due to proximity effects with metallic substrates. This extrinsic effect is momentum independent in the single valley approximation, and causes an energy splitting between opposite spin states of $13-225\,{\rm meV}$ \cite{spin-split1,spin-split2,spin-split3,spin-split4}: three to four orders of magnitude greater than the splitting associated with graphene's intrinsic spin-orbit coupling\cite{GSOC1,GSOC2}. Such a substantial enhancement of the RSO interaction has been mainly attributed to hybridization between carbon's $2p_{z}$ orbitals and the metal substrate's $d$ orbitals, and broken lattice symmetry in graphene\cite{excha}.  Moreover, the RSO interaction strength has been shown to be tunable via external gate voltages, as well as local lattice deformations\cite{curvedgraphne,Berche.2017}.

\begin{figure*}[t!]
    \centering
   \includegraphics{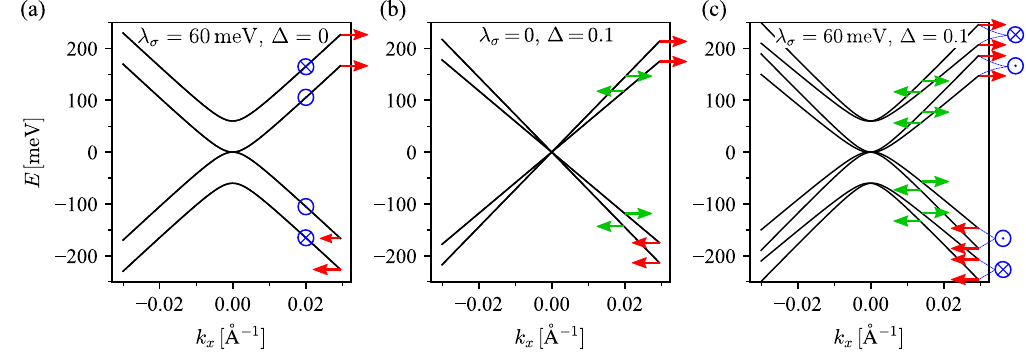}
    \caption{Band structure \eqref{eq:KekYBands} for Kek-Y graphene, for parameters (a) $\lambda_\sigma\ne0$ and $\Delta=0$; (b) $\lambda_\sigma=0$ and $\Delta\ne0$; and (c) $\lambda_\sigma\ne0$ and $\Delta\ne0$. In all cases, the Kekul\'e-indiced modulation to the RSO coupling was set to $\xi=0$. Blue, red and green arrows indicate the spin, sublattice and valley polarizations, respectively, with $\otimes$ and $\odot$ representing arrows pointing in the negative and positive $\hat{\mathbf{z}}$ directions, respectively.}
    \label{fig:KekYBands}
\end{figure*}
Given the bond length dependency of the Rashba parameter\cite{Berche.2017}, the following question naturally arises: How will the RSO coupling be modified by the presence of Kekulé lattice distortions in graphene? In this paper, we aim to answer this question. We introduce a generalized tight-binding Hamiltonian for graphene with RSO coupling and Kekulé lattice distortion, parameterized via position-dependent hopping- and spin-orbit interaction terms, exhibiting the Kekul\'e periodicity. After mapping the problem onto reciprocal space and folding the graphene bands onto the Kekul\'e Brillouin zone (KBZ), we derive effective $\mathbf{k}\cdot\mathbf{p}$-type Hamiltonians for both the Kek-Y and Kek-O textures, valid for the bands nearest the Fermi level. We then use these effective Hamiltonians to compute the low-energy band spectra of Kekul\'e distorted graphene with RSO coupling, and discuss its most salient features, including the resulting spin, pseudospin and valley textures. Finally, we introduce an out-of-plane magnetic field in the minimal coupling approximation\cite{luttinger_kohn} for the Kek-Y case, and compute its Landau level (LL) spectrum, focusing on the competing effects of the Kekul\'e and RSO terms in magnetotransport. In particular, we identify different trends for the horizontal LL splittings in a carrier density vs.\ magnetic field diagram, in the cases of graphene with only a Kek-Y distortion, graphene with only RSO coupling, and graphene with both a Kek-Y distortion and RSO coupling. We propose that these distinct trends may be used to experimentally identify the presence of Kek-Y distortions, RSO coupling, or both, on graphene-metal hybrid structures.

\section{Tight-binding model}

\subsection{Tight-binding model}\label{tbmodel}
The tight-binding Hamiltonian for a single layer of graphene with proximity-induced, sublattice-resolved Dirac-Rashba spin-orbit coupling\cite{kanemele1,kanemele2}, and a Kekul\'e lattice distortion\cite{Gamayun_2018}, can be written as $H^{pq} =H_{0}^{pq}+H_{R}^{pq}$, with the spinless graphene Hamiltonian
\begin{equation}\label{H0}
 H_0^{pq} = -\sum_{\mu}\sum_{\RR}\sum_{j=1}^3 \left(t_{\RR,\RR+\ddelta_j}^{pq}a^{\dagger}_{\RR,\mu}b_{\RR+\ddelta_j,\mu}+ {\rm H.c.}\right),
\end{equation}
where $a_{\RR,\mu}^\dagger$ ($a_{\RR,\mu}$) is the  creation (annihilation) operator
for an electron on site $\RR$ of sublattice $A$, with spin projection $\mu=\uparrow,\downarrow$, and $b_{\RR+\ddelta_j,\mu}^\dagger$ ($b_{\RR+\ddelta_j,\mu}$) are the corresponding $B$ sublattice operators. The integers $p,\,q$ parametrize the Kekul\'e bond texture over the honeycomb lattice, as shown below [see Eqs.\ \eqref{tlj} and \eqref{tR}]. In the case of pristine graphene $(\Delta=0)$, each atom at site $\RR$ is connected with three nearest neighbors at sites $\RR+\ddelta_j$, with relative position vectors  $\pmb{\delta}_{1} = \frac{a_0}{2}( \sqrt{3},-1 ),\
\pmb{\delta}_{2} = -\frac{a_0}{2}( \sqrt{3},1 )$ and $
\pmb{\delta}_{3} = a_0(0,1)$, where $a_{0}=1.421$\AA\, is the unperturbed C-C bond length. The lattice vectors are $
\mathbf{a}_{1} = \ddelta_{3}-\ddelta_{1}$ and $
\mathbf{a}_{2} = \ddelta_{3}-\delta_{2}$. However, the presence of a Kekul\'e bond distortion will modulate the hopping terms $t_{\RR,\RR+\ddelta_j}^{pq}$ as\cite{Gamayun_2018} 
 \begin{equation}\label{tlj}
 \frac{t_{\RR,\RR+\ddelta_j}^{pq}}{t_0} =1 +2{\rm Re}\left[ \Delta e^{i\KK_{pq}\cdot \pmb{\delta}_j+i\textbf{G}\cdot\RR}\right].
 \end{equation}
Here, $\KK_{pq}\equiv p\KK_+ + q\KK_-$, with the graphene valley vectors
\begin{equation}
    \KK_\pm = \frac{4\pi}{3\sqrt{3}a_0}\left(\pm\frac{1}{2},\,\frac{\sqrt{3}}{2} \right),
\end{equation}
and the vector
\begin{equation}
\GG\equiv\KK_{+}- \KK_{-}=\frac{4 \pi}{3\sqrt{3}a_0} (1,0)
\end{equation}
is a Kekul\'e superlattice primitive Bragg vector. The type of bond texture is determined by the integer ($\in \mathbb{Z}_{3}$) 
\begin{equation}\label{eq:nkek}
    n=(1+q-p)\, \mathrm{mod} 3,
\end{equation}
where a Kek-O texture corresponds to $n=0$, whereas Kek-Y textures are obtained for $n =\pm 1$.  
\begin{figure*}[t!]
    \centering
   \includegraphics[width=2\columnwidth]{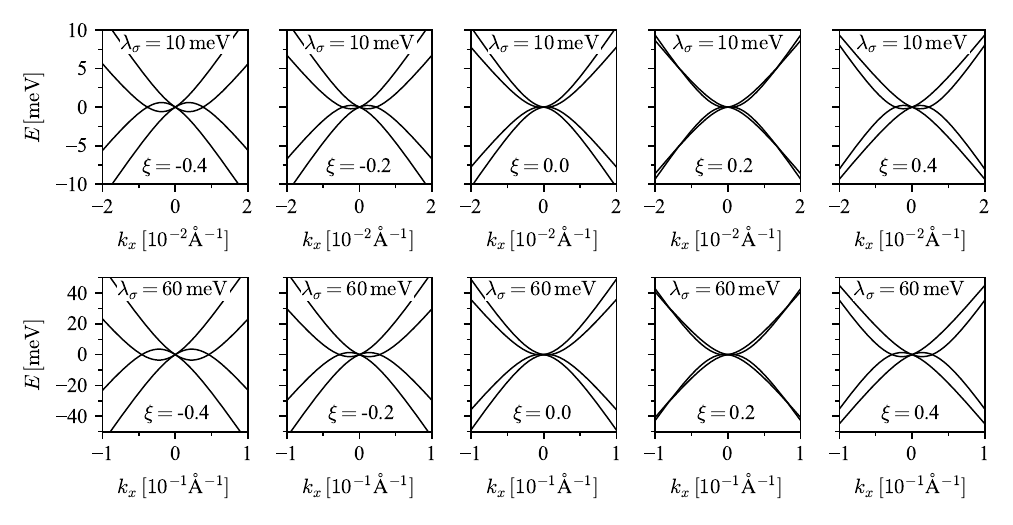}
    \caption{Low energy bands of Kek-Y graphene for different values of the Kekul\'e-induced RSO modulation $\xi$, for fixed $\Delta=0.1$, and $\lambda_\sigma=10\,{\rm meV}$ (top row) and $\lambda_\sigma=60\,{\rm meV}$ (bottom row). Note the qualitative similarities between the two cases when both axes are appropriately rescaled.}
    \label{fig:KekYBands_xi}
\end{figure*}

Since $\GG$ connects the two graphene valleys $\KK_\pm$ in reciprocal space, the latter are folded down onto the $\Gamma$-point of the KBZ. The hopping term modulation amplitude $\Delta$ is, in general, complex valued, although in the remainder of this paper we shall take both $\Delta$ and $t$ as real, without loss of generality. Finally, we point out that, for pristine graphene, the nearest-neighbor hopping integral reduces to $t_0=3.16\,{\rm eV}$.

The Dirac-Rashba spin-orbit term has the form
\begin{equation}\label{HR}
H_{R}^{pq}=\sum_{\mu\nu}\sum_{\RR}\sum_{j=1}^3\left[ \frac{i\lambda_{\RR,\RR+\ddelta_j}^{pq}}{2}a^{\dagger}_{\RR,\mu}(\vs_{\mu\nu} \times {\hat{\pmb{\delta}}}_{j} )_{z}b_{\RR+\ddelta_j,\nu}- {\rm H.c.}\right],
\end{equation}
where $\vs$ is the vector of Pauli matrices acting on the physical spin subspace, whereas $\hat{\ddelta}_j=\ddelta_j/|\ddelta_j|$. We have allowed a periodic bond-length modulation of the Rashba spin-orbit coupling $\lambda_{\RR,\RR+\ddelta_j}^{pq}$, analogous to that of the hopping parameter in Eq.\ (\ref{tlj}), with a complex amplitude $\xi$:
\begin{equation}\label{tR}
 \frac{\lambda_{\RR,\RR+\ddelta_j}^{pq}}{\lambda_{R}} =1 +2\mathrm{Re}\left[ \xi  e^{i\KK_{pq}\cdot \ddelta_j+i\GG\cdot\RR}\right],
 \end{equation}
where $\lambda_{R}$ is the Dirac-Rashba parameter in the absence of a Kekul\'e distortion.

\subsection{Total Hamiltonian in reciprocal space}
Taking the Fourier transforms of the total Hamiltonian gives the spin-conserving and Dirac-Rashba terms as
\begin{widetext}
\begin{equation}\label{eq:Hk}
\begin{split}
H_{0}^{pq} =&  \sum_{\kk\in{\rm BZ}}\sum_\mu\Big[   \Phi(\kk) a^{\dagger}_{\kk,\mu}   b_{\kk,\mu}
+ \Delta \Phi(\kk+\KK_{pq} ) a^{\dagger}_{\kk+\GG\, \mu} b_{\kk,\mu} +  \Delta \Phi(\kk-\KK_{pq}) a^{\dagger}_{\kk-\GG\,\mu} b_{\kk,\mu} +{\rm H.c.} \Big],\\
H_{R}^{pq} =& i\sum_{\kk\in{\rm BZ}}\sum_{\mu,\nu}\Big[a_{\kk,\mu}^\dagger \Lambda_{\mu\nu}(\kk) b_{\kk,\nu} + \xi a_{\kk+\GG,\mu}^\dagger \Lambda_{\mu\nu}(\kk+\KK_{pq}) b_{\kk,\nu} + \xi a_{\kk-\GG,\mu}^\dagger \Lambda_{\mu\nu}(\kk-\KK_{pq}) b_{\kk,\nu} - \mathrm{H.c}\Big],
\end{split}
\end{equation}
with the sum over $\kk$ running over all wave vectors of the original (pristine graphene) Brillouin zone (BZ). We have also defined
\begin{equation}\label{eq:phiandlambda}
\Phi(\kk)=-t_0\sum_{j=1}^3e^{i\kk\cdot\ddelta_j},\quad   
\Lambda_{\mu\nu}(\kk) = \frac{\lambda_R}{2}\sum_{j=1}^3e^{i\kk\cdot\ddelta_j}\,(\vs_{\mu\nu}\times\hat{\ddelta}_j)_z.
\end{equation}
\end{widetext}

\begin{figure}[t!]
    \centering
    \includegraphics{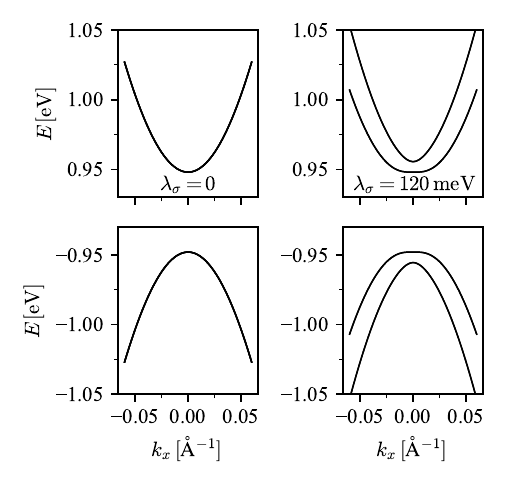}
    \caption{Band structures of the Hamiltonian \eqref{eq:keko} for Kek-O graphene, for RSO coupling $\lambda_\sigma=0$ and $120\,{\rm meV}$, for fixed $\Delta=0.1$.}
    \label{fig:KekOBands}
\end{figure}

We now perform a zone folding from the original BZ onto the KBZ by introducing the 12-spinors $\Psi_{\kk}=(c_{\kk,\uparrow},\,c_{\kk,\downarrow})^T$, where
 \begin{equation}\label{eq:cdef}
c_{\kk,\mu}=\left( a_{\kk,\mu},\,a_{\kk-\GG,\mu},\,a_{\kk+\GG,\mu},\,b_{\kk-\GG,\mu},\, b_{\kk+\GG,\mu},\,b_{\kk,\mu}\right)^T,
\end{equation}
and $\kk\in {\rm KBZ}$. Using the properties ($m\in\mathbb{Z}$)
\begin{equation}
\Xi (\kk)=e^{i \frac{2\pi}{3}m}\Xi (\kk+ m(\KK_+ +\KK_-))=\Xi (\kk+3m\KK_{\pm}),
\end{equation}
valid for both $\Xi (\kk)=\Phi(\kk)$ and $\Xi(\kk) = \Lambda(\kk)$, we may write
\begin{equation}
H^{pq}= \sum_{\kk\in{\rm KBZ}} \Psi^{\dagger}_{\kk}\Hcal_{pq}(\kk)\Psi_{\kk},
\label{H}
\end{equation}
\begin{figure*}[t!]
    \centering
    \includegraphics{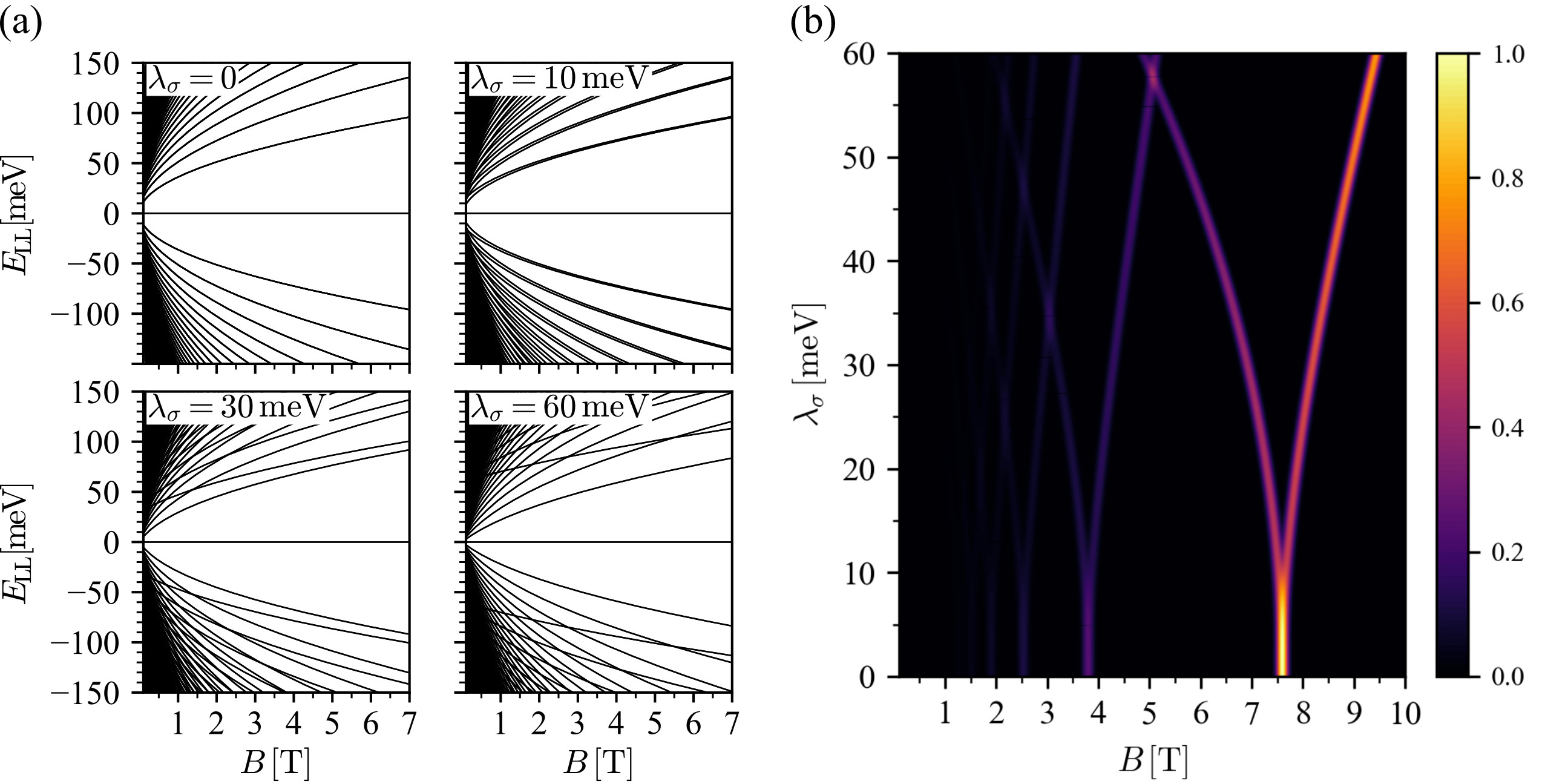}
    \caption{(a) Landau levels for different values of $\lambda_\sigma$, and (b) $B^2\cdot {\rm DOS}(\varepsilon_F)$ (arbitrary units) as a funcion of $\lambda_\sigma$ and magnetic field, for $\Delta=\xi=0$. In panel (b) we have set $\varepsilon_F=50\,{\rm meV}$.}
    \label{fig:LLJustRashba}
\end{figure*}
with the Bloch Hamiltonian
\begin{equation}
\Hcal_{pq}(\kk) = \begin{pmatrix}
    0 & \Ecal_n(\kk)         & 0 & i\Rcal_{n}^{\uparrow\downarrow}(\kk) \\
    \Ecal_n^\dagger(\kk) & 0 & -i\Rcal_{n}^{\downarrow\uparrow\,*}(\kk) & 0 \\
    0 & i\Rcal_{n}^{\downarrow\uparrow}(\kk) & 0 & \Ecal_n(\kk)     \\
    -i\Rcal_{n}^{\uparrow\downarrow}{}^*(\kk) & 0 & \Ecal^\dagger_n(\kk) & 0
\end{pmatrix}
\end{equation}
containing the matrices
\begin{equation}
\begin{split}
\Ecal_{n}(\kk) =&
\begin{pmatrix}
    \Phi_0(\kk) & \Delta_{pq}\Phi_{n+1}(\kk) & \Delta_{pq}^* \Phi_{-n-1}(\kk) \\
    \Delta_{pq}^*\Phi_{-n+1}(\kk) & \Phi_{-1}(\kk) & \Delta_{pq}\Phi_n(\kk) \\
    \Delta_{pq}\Phi_{n-1}(\kk) & \Delta_{pq}^*\Phi_{-n}(\kk) & \Phi_{1}(\kk)
\end{pmatrix},\\
\Rcal_n^{\mu\nu}(\kk)=&\begin{pmatrix}
    \Lambda_{0}^{\mu\nu}(\kk) & \xi_{pq} \Lambda_{n+1}^{\mu\nu}(\kk) & \xi_{pq}^*\Lambda_{-n-1}^{\mu\nu}(\kk)\\
    \xi_{pq}^*\Lambda_{-n+1}^{\mu\nu}(\kk) & \Lambda_{-1}^{\mu\nu}(\kk) & \xi_{pq} \Lambda_{n}^{\mu\nu}(\kk) \\
    \xi_{pq} \Lambda_{n-1}^{\mu\nu}(\kk) & \xi_{pq}^* \Lambda_{-n}^{\mu\nu}(\kk) & \Lambda_{1}^{\mu\nu}(\kk),
\end{pmatrix},
\end{split}
\end{equation}
where we have introduced the folded tunneling function $\Phi_n(\kk)\equiv \Phi(\kk+n\GG)$ and RSO coupling $\Lambda_n^{\mu\nu}(\kk)\equiv \Lambda_{\mu\nu}(\kk+n\GG)$, with $n$ given by Eq.\ \eqref{eq:nkek}. We have also introduced the the Kekul\'e terms
\begin{equation}
\Delta_{pq}\equiv e^{i\tfrac{2\pi}{3}(p+q)}\Delta,\quad \xi_{pq}\equiv e^{i\tfrac{2\pi}{3}(p+q)}\xi.
\end{equation}
Note that all terms $\Lambda_{\uparrow\uparrow}(\kk+n\GG)=\Lambda_{\downarrow\downarrow}(\kk+n\GG)=0$, by the symmetry of the Pauli matrices.

We may now obtain an effective low-energy model with reduced dimensionality $8\times 8$ by projecting out  the two high-energy electron- and hole bands present in $\Hcal_{pq}(\kk)$ for  $\kk$ near the $\Gamma$ point, corresponding to operators $a_{\kk,\mu}$ and $b_{\kk,\mu}$ in Eq.\ \eqref{eq:cdef}. We do so at zeroth order in perturbation theory, and linearize $\Phi_n(\kk)$ and $\Lambda_n(\kk)$ about $\kk=\boldsymbol{0}$. Finally, introducing the new $8$-spinor basis
\begin{equation}\label{eq:thisbasis}
\begin{split}
\Psi_{\kk}'=(-&b_{\kk-\GG\,\uparrow}, -b_{\kk-\GG\,\downarrow}, a_{\kk-\GG\,\uparrow}, 
a_{\kk-\GG\,\downarrow}, \\
&a_{\kk+\GG\,\uparrow},\,\,\,
a_{\kk+\GG\,\downarrow}, 
b_{\kk+\GG\,\uparrow},
b_{\kk+\GG\,\downarrow})^T,
\end{split}
\end{equation}
we obtain two compact forms of $\tilde{\Hcal}_{pq}$ (Appendix  \ref{app:Hkp}): all cases when $(1+q-p){\rm mod}3=\pm1$ give the Kek-Y Bloch Hamiltonians
\begin{equation}\label{eq:keky}
\begin{split}    
    \Hcal_{\rm Y}(\kk)=&\hbar v_\sigma \tau_0(\kk\cdot\ssigma)s_0 + \frac{\lambda_\sigma}{2} \tau_0(\ssigma\times \vs)_z\\ &+ \hbar v_\tau (\kk\cdot\ttau_\pm)\sigma_0s_0+ \frac{\lambda_\tau}{2} (\ttau_\pm\times\sigma_0\mathbf{s})_z,
\end{split}
\end{equation}
whereas for $(1+q-p){\rm mod}3=0$ we obtain the Kek-O model
\begin{equation}\label{eq:keko}
    \Hcal_{\rm O}(\kk)=\hbar v_\sigma \tau_0(\kk\cdot\ssigma)s_0 + \frac{\lambda_\sigma}{2} \tau_0(\ssigma\times\vs)_z + 3t_0\Delta \tau_x\sigma_zs_0.
\end{equation}
Here, we have used the standard definitions of the Pauli matrix vectors $\ssigma$ and $\vs$ acting on the sublattice and spin degrees of freedom, respectively. For the valley subspace, we have defined the Pauli vectors $\ttau_n = (n\tau_x,\tau_y,\tau_z)$, where $n=\pm1$ corresponds to the type of Kek-Y texture defined by the parameters $p$ and $q$.  $\tau_0$, $\sigma_0$ and $s_0$ are the unit matrices in the valley, sublattice and spin subspaces. We have also defined the two Fermi velocities $v_\sigma = \tfrac{3}{2\hbar}t_0a_0$ and $v_\tau = \Delta v_\sigma$, and the constants $\lambda_\sigma=\tfrac{3}{2}\lambda_R$ and $\lambda_\tau=\xi\lambda_\sigma$. The latter, introduced by the Kekul\'e-modulated RSO interaction in the Kek-Y case, constitutes a novel spin-valley coupling. By contrast, note that in a Kek-O texture there is no coupling between valley and momentum, or valley and spin. 

Both models $\Hcal_{\rm Y}$ and $\Hcal_{\rm O}$ can be diagonalized exactly, yielding the band structures
\begin{subequations}
\begin{equation}\label{eq:KekYBands}
\begin{split}
    E^{\rm Y}_{\alpha,\beta,\gamma}(\kk) =& \frac{\alpha}{2}\Big[\sqrt{(2\hbar v_\tau k + \beta\gamma \lambda_\sigma)^2+\lambda_\tau^2}\\
    &+ \gamma\sqrt{(2\hbar v_\sigma k + \beta\gamma \lambda_\tau)^2+\lambda_\sigma^2}\Big],
\end{split}
\end{equation}
\begin{equation}\label{eq:KekOBands}
\begin{split}
    E^{\rm O}_{\alpha,\beta}(\kk) =& \alpha\Big[(\hbar v_\sigma k)^2 + (3t_0\Delta)^2+\tfrac{\lambda_\sigma^2}{8}\\
    &+\beta\lambda_\sigma\sqrt{(\hbar v_\sigma k)^2+(\tfrac{\lambda_\sigma}{2})^2}\Big]^{1/2},
\end{split}
\end{equation}
\end{subequations}
where the indices $\alpha,\beta,\gamma=\pm1$. Note that in the Kek-Y case, the dispersions are identical for $n=\pm1$.  Figure \ref{fig:KekYBands} shows the band structure \eqref{eq:KekYBands} along the $\overline{-M\,\Gamma\,M}$ line ($k_y=0$) of the KBZ, for different values of the RSO coupling $\lambda_\sigma$ and Kekul\'e hopping modulation $\Delta$, keeping the Rashba modulation parameter $\xi=0$. Setting $\Delta=0$ for finite $\lambda_\sigma$, we obtain the well known band structure of RSO-coupled graphene\cite{PhysRevB.79.161409,spin-split1}, except folded onto the KBZ, leading to a double degeneracy for each band, corresponding to the valley pseudo-spin. Figure \ref{fig:KekYBands}(a) also shows the expectation values of the $\vs$ and $\ssigma$ operators as blue and red arrows, respectively, showing that the sublattice polarization for all bands is locked perpendicularly to the spin, forming a right-handed (left-handed) pair for the first conduction and second valence (second conduction and first valence) bands, with the sublattice (spin) vector always pointing in the radial (polar) direction.

Figure \ref{fig:KekYBands}(b) shows the case of $\lambda_\sigma=0$ with a finite Kekul\'e hopping modulation $\Delta$, reproducing the band structure of ordinary Kek-Y graphene\cite{Gamayun_2018}, consisting of two concentric Dirac cones with different Fermi velocities, $v_\sigma \pm v_\tau$. All bands are spin degenerate and valley-sublattice locked into parallel (second conduction- and valence bands) or anti-parallel (first conduction- and valence bands) pairs, with both vectors always oriented radially. This is shown in Fig.\ \ref{fig:KekYBands}(b), where the sublattice and valley vectors are shown with red and green arrows, respectively.

Next, Fig.\ \ref{fig:KekYBands}(c) shows the band structure for Kek-Y graphene ($\Delta\ne0$) with a finite RSO coupling ($\lambda_\sigma\ne0$). As in the case of ordinary Kek-Y graphene, each band has radial valley and sublattice polarizations, locked into parallel or anti-parallel pairs. Moreover, for each band the spin orientation is also locked with the sublattice vector, forming either left- or right-handed pairs. All band structures if Fig.\ \ref{fig:KekYBands} are particle-hole symmetric, as a consequence of the chiral symmetry $\{ \Hcal_{\rm Y}(\kk),\,\tau_z\sigma_zs_0\}=0$, which is exact for all parameter values. This chirality operator was first identified by Gamayun \emph{et al}.\cite{Gamayun_2018} for ordinary Kek-Y graphene ($\lambda_\sigma=0$).

Finally, Fig.\ \ref{fig:KekYBands_xi} shows how the Kekul\'e-induced modulation to the RSO coupling $\xi$ modifies the band structure of Kek-Y graphene, focusing on the first two conduction- and valence bands, and choosing large values $|\xi|=0.2,\,0.4$ to clearly see its effects on the band structure. For either positive or negative $\xi$, Fig.\ \ref{fig:KekYBands_xi} reveals the appearance of a doubly degenerate Dirac cone centered at the $\Gamma$ point, surrounded by a circular band touching at the Fermi level centered at the $\Gamma$ point, with bands that disperse linearly in the radial direction away from the touching points. The resulting Fermi surface is a nondegenerate circle surrounding a doubly degenerate point at $\Gamma$. The radius of the Fermi circle increases with $|\xi|$, and the case $\xi=0$ represents a critical point where the Fermi surface becomes a quadruply degenerate point at $\Gamma$.

For completeness, Fig.\ \ref{fig:KekOBands} shows the band structures of Kek-O graphene for $\lambda_\sigma=0$ and $60\,{\rm meV}$, and fixed $\Delta=0.1$. It exhibits a large, direct band gap of size $6t_0\Delta \approx 1.9\,{\rm eV}$ at the $\Gamma$ point, coming from the valley-sublattice coupling in Eq.\ \eqref{eq:keko}. For $\lambda_\sigma = 0$, the conduction and valence bands are parabolic, whereas for finite $\lambda$ we obtain the typical band structure of RSO-coupled parabolic bands. Note that the the Kekul\'e-induced Rashba modulation $\xi$ does not appear in the Kek-O Hamiltonian \eqref{eq:keko}. Henceforth, we shall focus on the Kek-Y case, which we deem more interesting due to its lack of a band gap and the chiral nature of its bands.

\begin{figure*}[t!]
    \centering
    \includegraphics{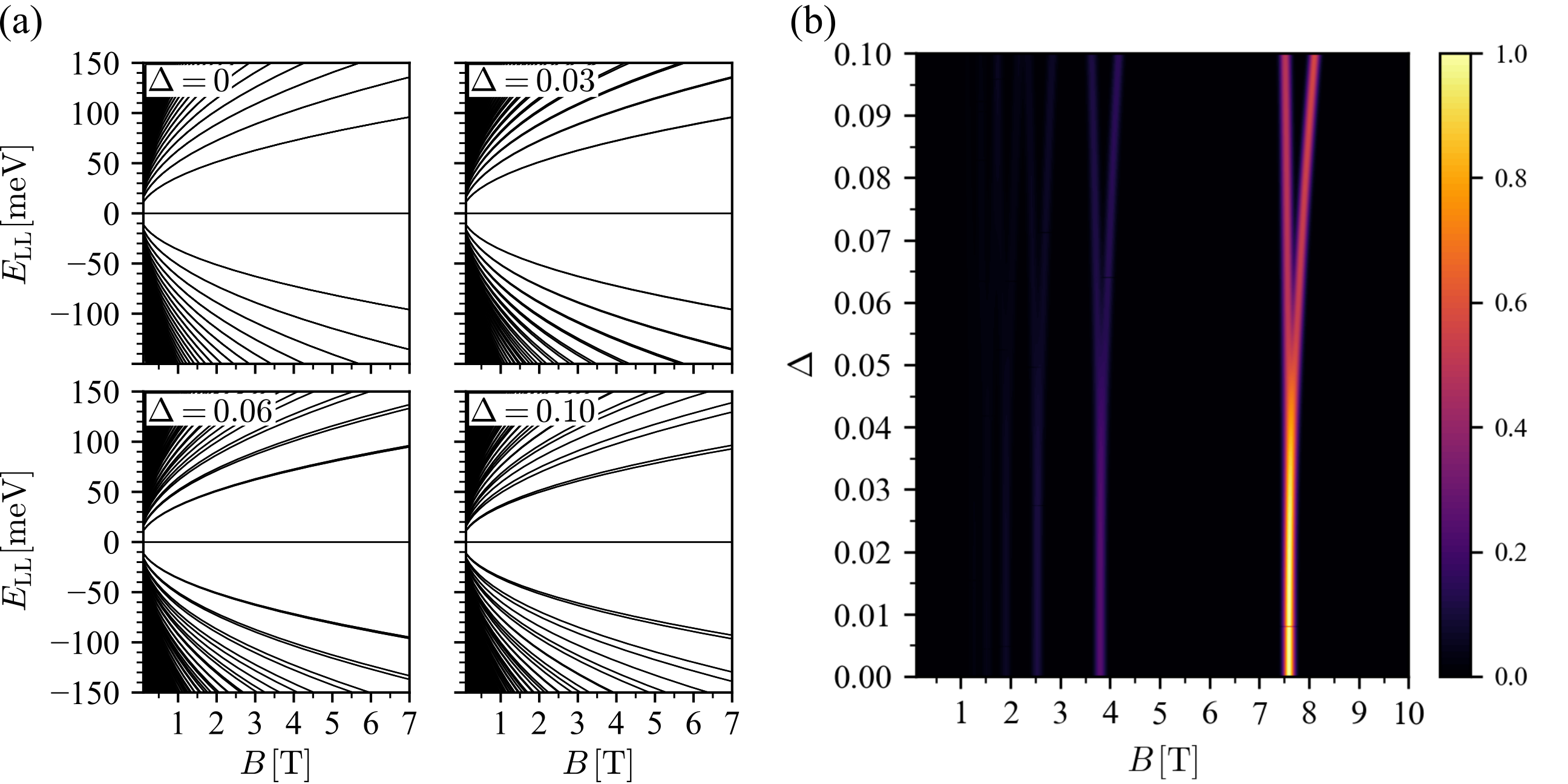}
    \caption{(a) Landau levels for different values of $\Delta$, and (b) $B^2\cdot {\rm DOS}(\varepsilon_F)$ (arbitrary units) as a funcion of $\Delta$ and magnetic field, for $\lambda_\sigma=\xi=0$. In panel (b) we have set $\varepsilon_F=50\,{\rm meV}$.}
    \label{fig:LLJustKek}
\end{figure*}
\section{Landau level spectrum of Kek-Y graphene with RSO coupling}\label{sec:LLs}
We introduce an out-of-plane magnetic field $\mathbf{B}=B \hat{\mathbf{z}}$, with symmetric-gauge vector potential $\mathbf{A} = \tfrac{B}{2}(-y\hat{\mathbf{x}}+x\hat{\mathbf{y}})$, into the Kek-Y graphene Hamiltonian \eqref{eq:keky}\cite{Mohammadi_2022}, via the minimal substitution $\hbar \kk \rightarrow \hbar \kk - \tfrac{e}{c} \mathbf{A}=\ppi$, where the components of the canonical momentum $\ppi$ obey the algebra $[\pi_x,\,\pi_y]=-i\tfrac{e}{c}\hbar B$. This allows the introduction of the ladder operators ($\pi_\pm \equiv \pi_x \pm i \pi_y$)
\begin{equation}
a = \sqrt{\frac{c}{2e\hbar B}}\pi_-,\quad a^\dagger = \sqrt{\frac{c}{2e\hbar B}}\pi_+,
\end{equation}
obeying the harmonic oscillator algebra $[a,a^\dagger] = 1$, and operating on the Landau level (LL) basis $\{| \ell \rangle\}$ as $a| \ell \rangle=\sqrt{\ell}| \ell-1 \rangle$ and $a^\dagger | \ell\rangle=\sqrt{\ell+1}|\ell+1 \rangle$, for integer $\ell\ge0$. These operators enter the Kek-Y Hamiltonian \eqref{eq:keky} through the substitution 
\begin{equation}
    \ppi \cdot \boldsymbol{\nu} = \pi_+ \nu_- + \pi_- \nu_+ = \sqrt{\frac{2e\hbar B}{c}}\left( a^\dagger \nu_- + a \nu_+ \right),
\end{equation}
where $\boldsymbol{\nu}=\ttau,\,\ssigma$. The resulting LL Hamiltonian is shown in its full form in Appendix \ref{app:LLHamiltonian}. Here, we merely report its numerical energy spectra for varying magnetic field.

Figure \ref{fig:LLJustRashba}(a) shows the LLs obtained in the absence of a Kek-Y deformation, for multiple values of the RSO coupling $\lambda_\sigma$, including $\lambda_\sigma=0$, which corresponds to the case of pristine graphene. In that case \cite{mcclure1956diamagnetism}, a four-fold degenerate zero energy mode appears for all magnetic field values, surrounded by an electron-hole symmetric fan of valley- and spin-degenerate LLs evolving as $B^{1/2}$. The zero modes persist for finite $\lambda_\sigma$, and the surrounding LLs split into two distinct fans, corresponding to the two separate conduction- and valence bands shown in Fig.\ \ref{fig:KekYBands}(a). As a connection with transport experiments, Fig.\ \ref{fig:LLJustRashba}(b) shows the density of states (DOS) at the Fermi level as a function of both magnetic field and $\lambda_\sigma$, setting $\varepsilon_F=50\,{\rm meV}$.

Next, Fig.\ \ref{fig:LLJustKek} shows the LL spectrum of Kek-Y graphene for multple values of $\Delta$, setting $\lambda_\sigma=\xi=0$, showing that the zero-mode cuadruplet also survives in the presence of the Kekul\'e deformation, as reported in Ref.\ \onlinecite{Gamayun_2018}. The reference LL fan of pristine graphene splits into two separate fans when $\Delta\ne0$. Note, however, that this splitting occurs only at finite magnetic fields, and both fans evolve with magnetic field as $B^{1/2}$, by contrast to the case of Rashba-SO-coupled graphene. This is a consequence of the two Dirac cones with different Fermi velocities $v_\sigma$ and $v_\tau$, shown in Fig.\ \ref{fig:KekYBands}(b). Figure \ref{fig:LLJustKek}(b) shows the Fermi-level DOS for $\varepsilon_F=50\,{\rm meV}$, as it evolves with the Kekul\'e hopping modulation $\Delta$. Importantly, we can see that the DOS peaks split with increasing $\Delta$, analogously to the case of RSO coupling shown in Fig.\ \ref{fig:LLJustRashba}(b). In other words, a splitting in the DOS peaks, measured in magnetotransport experiments as split conductance peaks, may come from either source. However, as we discuss next, the two effects can be distinguished through doping dependent transport measurements. For completeness, the LL spectrum of graphene with both a Kek-Y distortion and RSO coupling is shown in Appendix \ref{app:LLsKekYandRSO}.

\begin{figure*}[t!]
    \centering
    \includegraphics{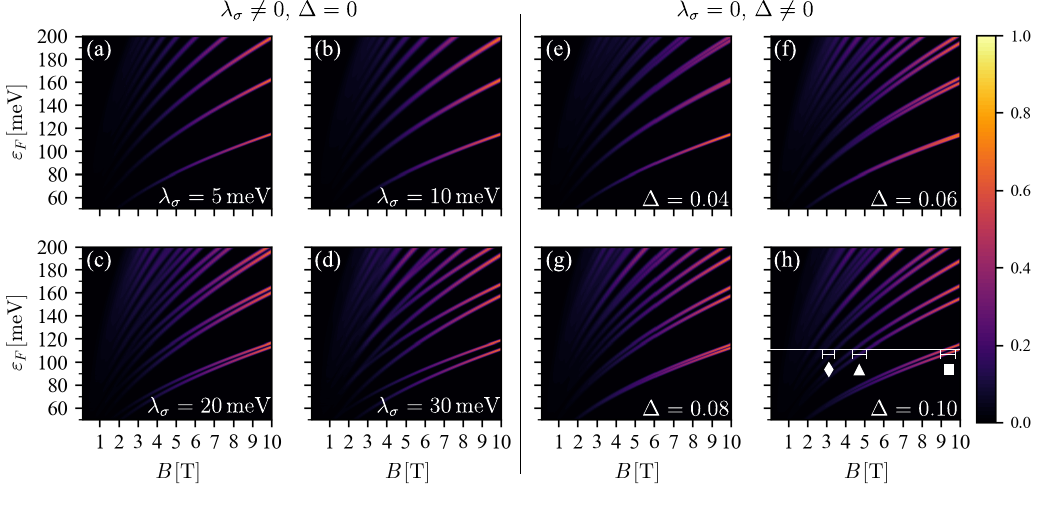}
    \caption{$B^2\cdot {\rm DOS}(\varepsilon_F)$ as a funcion of the Fermi energy $\varepsilon_F$ and magnetic field $B$, for $\xi=0$. Left (right) panels correspond to model \eqref{eq:keky} with $\Delta=0$ ($\Delta=0$). The markers in panel (h) indicate the splittings described in Fig.\ \ref{fig:splittings}.}
    \label{fig:DOSvsEf}
\end{figure*}
Figure \ref{fig:DOSvsEf} shows the DOS at the Fermi level for varying magnetic field $B$ and Fermi energy $\varepsilon_F$, keeping the model parameters in \eqref{eq:keky} constant, with $\xi=0$ in all cases. Qualitatively similar behaviors are observed both for finite $\lambda_\sigma$ and no Kek-Y distortion (left panels) and for no RSO coupling with finite Kek-Y distortion (right panels); namely, a fan of LLs that split into two, with overall larger splittings obtained for larger values of the finite parameter. Some of these splittings are indicated in Fig.\ \ref{fig:DOSvsEf}(h). However, we have found that the $\varepsilon_F$-dependence of these splittings are distinct for the cases of $(\lambda_\sigma\ne0,\Delta=0)$ and $(\lambda_\sigma=0,\Delta\ne0)$. 

\begin{figure*}[t!]
    \centering
    \includegraphics{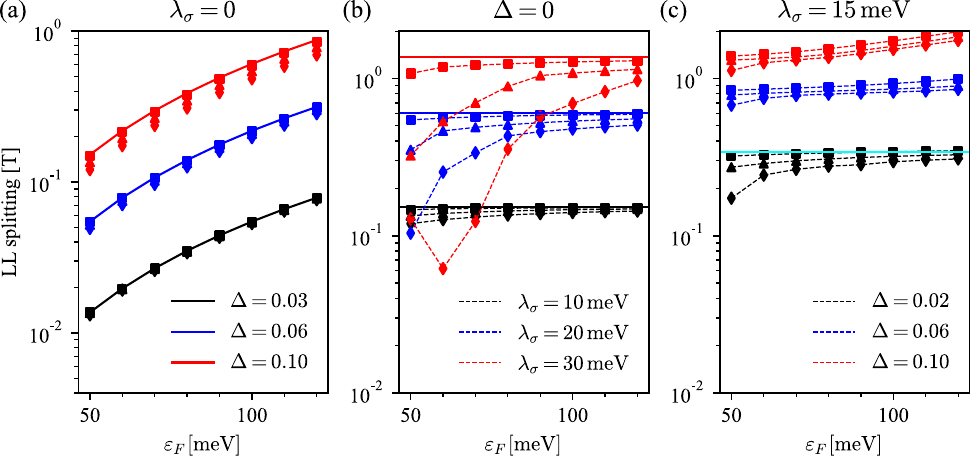}
    \caption{Energy splittings of the lower three LL pairs as functions of the Fermi energy $\varepsilon_F$, for fixed model parameters (a) $(\lambda_\sigma=0,\,\Delta\ne0)$, (b) $(\lambda_\sigma\ne0,\,\Delta=0)$ and (c) $(\lambda_\sigma\ne0,\,\Delta\ne0)$, with $\xi=0$ in all cases. The data symbols ($\blacksquare$, $\blacktriangle$ and $\blacklozenge$) are matched to those shown in Fig.\ \ref{fig:DOSvsEf}(h). The solid lines in panel (a) correspond to Eq.\ \eqref{eq:powerlaw} with the coefficient $A_\blacksquare$ given by Eq.\ \eqref{eq:Acoeff}, whereas in panel (b) the solid lines indicate the saturation value given by Eq.\ \eqref{eq:saturation}. The cyan solid line in panel (c) indicates the saturation value predicted by Eq.\ \eqref{eq:saturation}, which is exceeded in the presence of a Kek-Y distortion. The dashed lines in panels (b) and (c) are merely guides to the eye.}
    \label{fig:splittings}
\end{figure*}

Figure \ref{fig:splittings} shows the LL magnetic-field splittings of the bottom three split pairs indicated in Fig.\ \ref{fig:DOSvsEf}(h), as functions of the Fermi energy, setting $\lambda_\sigma=0$. Figure \ref{fig:splittings}(a), shows that, for a pure Kek-Y distortion ($\lambda_\sigma=0$), all three splittings exhibit a power-law behavior
\begin{equation}\label{eq:powerlaw}
    \delta B_{\rho}(\varepsilon_F)=A_\rho \, \varepsilon_F^2,
\end{equation}
where $\rho=\blacksquare,\,\blacktriangle,\,\blacklozenge$ indicates the corresponding magnetic field splitting shown in Fig.\ \ref{fig:DOSvsEf}. For instance, for the rightmost splitting one can analytically compute the coefficient (see Appendix \ref{app:asymptotic})
\begin{equation}\label{eq:Acoeff}
A_\blacksquare = \frac{8\Delta^2}{\hbar^2 v_F^2}\sqrt{\frac{c}{2e\hbar}},
\end{equation}
up to third order in $\Delta$. By contrast, Fig.\ \ref{fig:splittings}(b) shows that, in the case of pure RSO coupling ($\Delta=0$), there is a clear saturation of the splitting energies at large $\varepsilon_F$, with the first LL splitting (symbol $\blacksquare$) showing saturation already for $\varepsilon_F\approx 50\,{\rm meV}$. The saturation value of the first LL splitting can be computed as (see Appendix \ref{app:asymptotic})
\begin{equation}\label{eq:saturation}
    \delta B_{\blacksquare}(\varepsilon_F\gg\lambda_\sigma) = \frac{2\lambda_\sigma^2}{\hbar^2v_\sigma^2}\sqrt{\frac{c}{2e\hbar}}.
\end{equation}
We propose that these distinct behaviors may be used experimentally, not only to distinguish between the two effects, but also to estimate the magnitude of the Kekul\'e distortion or RSO coupling, using doping-dependent magnetotransport measurements.

Figure \ref{fig:splittings}(c) shows the predicted magnetic field splittings for a graphene sample with both a Kek-Y distortion, and a finite RSO coupling, the latter without a Kekul\'e modulation ($\xi=0$). Although the analysis that led us to Eqs.\ \eqref{eq:Acoeff} and \eqref{eq:saturation} can be repeated in this case, the resulting expressions are much more complicated and far less illuminating. Nonetheless, the simultaneous presence of both the Kek-Y distortion and the Rashba effect can be inferred from the first ($\blacksquare$) splitting: Figure \ref{fig:splittings}(c) shows a crossover from a saturating behavior at low $\varepsilon_F$, consistent with $\lambda_\sigma \ne 0$, to a monotonic increase at large $\varepsilon_F$ when the Kek-Y distortion begins to dominate. Importantly, note that the initial plateau that appears before the crossover exceeds the theoretical value for the case of only RSO coupling \eqref{eq:saturation}, shown by the cyan line in Fig.\ \ref{fig:splittings}(c). Therefore, when aiming to determine the values of the Hamiltonian parameters $\lambda_\sigma$ and $\Delta$ from magnetotransport experiments, it is important to explore beyond the weak doping regime, to avoid overestimating the value of $\lambda_\sigma$.

Finally, we briefly discuss the case where all Kek-Y model parameters $\lambda_\sigma$, $\Delta$ and $\xi$ are finite. The corresponding LL spectrum is shown in Fig.\ \ref{fig:LLKekRashba} of Appendix \ref{app:LLsKekYandRSO}. In this case, the LL structure is quite complex, and a doping-dependent magnetic-field-splitting analysis becomes intractable. However, the new dispersion has a salient feature that is absent when either $\Delta=0$ or $\lambda_\sigma=0$; namely, that the zero-energy quadruplet shown in Figs.\ \ref{fig:LLJustRashba}(a) and \ref{fig:LLJustKek}(a) splits for finite magnetic fields into a zero-energy doublet, closely surrounded by two electron-hole symmetric LLs. For small values of $|\Delta - \xi|$, the splitting between the latter two LLs takes the form ($v_\sigma(B) \equiv v_\sigma \sqrt{\frac{2e\hbar B}{c}}$, see Appendix \ref{app:LLHamiltonian})
\begin{equation}\label{eq:zeromodesplittingmain}
    \delta\varepsilon_{z'} = \frac{2\lambda_\sigma v_\sigma(B)}{\sqrt{(1+\Delta^2)v_\sigma^2(B) + (1+\xi^2)\lambda_\sigma^2}} |\Delta - \xi |,
\end{equation}
and is always finite, with the exception of the fine-tuned case $\Delta = \xi$. It is possible that this feature may be observed in magnetotransport experiments at very low doping levels, thus confirming that both a Kek-Y texture and a finite RSO coupling are present in the graphene system the presence of a Kekul\'e-induced modulation of the RSO coupling in Kek-Y graphene. Moreover, if the parameters $\Delta$ and $\lambda_\sigma$ are estimated from the previously discussed analysis of the magnetic-field splittings, they may be introduced into Eq.\ \eqref{eq:zeromodesplittingmain}, thus allowing for a rough estimation of the RSO modulation parameter $\xi$.

\section{Conclusions}
We have introduced generalized tight-binding Hamiltonians for graphene with Kekul\'e-Y and Kekul\'e-O bond textures, as well as Rashba spin-orbit coupling, which takes into account a possible modulation of the Rashba term by the Kekul\'e bond distortions. These models aim to describe epitaxial Kekul\'e graphene on transition metal surfaces, which have been shown to induce sizable Rashba spin-orbit couplings through proximity effects. We have also derived low-energy effective models based on the general tight-binding Hamitonians, following the scheme introduced by Gamayun \emph{et al.}\cite{Gamayun_2018}. We have found that, whereas the Kekul\'e-O effective model is independent of the Rashba term modulation, the Kekul\'e-Y model exhibits a new spin-valley locking term that can dramatically modify the Fermi surface of the system in charge neutrality.

We have also studied the Landau level spectrum of this system under a perpendicular magnetic field, and computed its density of states at the Fermi level with an aim to motivate magnetotransport experiments on Kekul\'e graphene systems on transition metal substrates. We have found that, whereas the Landau level spectra exhibit a degenerate quadruplet of zero-energy modes in graphene with either a Kekul\'e-Y distortion or Rashba spin-orbit coupling, this degeneracy is partially lifted when both effects are present, resulting on a degenerate doublet of zero-energy modes closely surrounded by two satellite Landau levels. Based on our findings, we have put forth an experimental method to distinguish the presence of the Kekul\'e-Y bond texture, of a Rashba spin-orbit term, or both, based on an analysis of the magnetic-field splittings of the DOS peaks, as functions of the Fermi energy. Finally, we have shown that, once the strengths of both effects are extracted from this analysis, they may be used to estimate the magnitude of the Kekul\'e-induced modulation of the Rashba spin-orbit coupling based on the magnitude of the zero-energy mode splitting, which may be experimentally resolved at low enough values of the Fermi energy.

\begin{acknowledgments}
D.A.R.T.~acknowledges funding from PAPIIT-DGAPA-UNAM through project IA106523. F.M. acknowledges funding from PAPIIT-DGAPA-UNAM through project IN113920.
\end{acknowledgments}

\appendix

\section{Effective low-energy Bloch Hamiltonians}\label{app:Hkp}
Starting from Eq.\ (\ref{H}), we identify the four high-energy bands corresponding to states at the $\Gamma$ point of the original BZ, and neglect them, which corresponds to projecting them out at zeroth order in perturbation theory. This is justified by the large offset between these states and the Fermi level ($\approx 3\,{\rm eV}$), as compared with all relevant model parameters, which fall in the 10 meV range. Collecting the annihilation operators in the column vector $\psi_{\kk}=( a_{\kk-\GG,\uparrow}$, $a_{\kk-\GG,\downarrow}$,$a_{\kk+\GG,\uparrow}$,$a_{\kk+\GG,\downarrow}$,$b_{\kk-\GG,\uparrow}$,$b_{\kk-\GG,\downarrow}$,$b_{\kk+\GG,\uparrow}$,$b_{\kk+\GG,\downarrow})^T$, we write the total Hamiltonian for the eight bands closest to the Fermi level as
\begin{equation}
    \Hcal_{pq}'(\kk)=\begin{pmatrix}
        0 & \Sigma_n(\kk) \\
        \Sigma^\dagger_n(\kk) & 0
    \end{pmatrix},
\end{equation}
with $n$ given by the Kekul\'e texture parameters $p$ and $q$ through Eq.\ \eqref{eq:nkek}, and
\begin{equation}
    \Sigma_n(\kk)=\begin{pmatrix}
        \Phi_{-1}(\kk) & i\Lambda_{-1}^{\uparrow\downarrow}(\kk) & \Delta \Phi_n(\kk) & i\xi \Lambda_n^{\uparrow\downarrow}(\kk)\\
        i\Lambda_{-1}^{\downarrow\uparrow}(\kk) & \Phi_{-1}(\kk) & i\xi\Lambda_n^{\downarrow\uparrow}(\kk) & \Delta\Phi_n(\kk) \\
        \Delta^*\Phi_{-n}(\kk) & i\xi^*\Lambda_{-n}^{\uparrow\downarrow}(\kk) & \Phi_1(\kk) & i\Lambda_1^{\uparrow\downarrow}(\kk)\\
        i\xi^*\Lambda_{-n}^{\downarrow\uparrow}(\kk) & \Delta^*\Phi_n(\kk) & i\Lambda_1^{\downarrow\uparrow}(\kk) & \Phi_1(\kk)
    \end{pmatrix}.
\end{equation}
where $\Phi_n(\kk)=\Phi_n(\kk+n\GG)$ and $\Lambda_n^{\mu\nu}(\kk)=\Lambda_{\mu\nu}(\kk+n\GG)$, following the definitions \eqref{eq:phiandlambda}. For $\mu\ne\nu$ we obtain explicitly
\begin{widetext}
\begin{subequations}
\begin{equation}
    \Lambda_0^{\mu\nu}(\kk)=\frac{\lambda_\sigma}{3}e^{ia_0k_y}\left[1+2\cos{\left(\tfrac{\sqrt{3}}{2}a_0k_x + \tfrac{2\pi}{3}\gamma_{\mu\nu} \right)e^{-i\tfrac{3}{2}a_0k_y}} \right],
\end{equation}
\begin{equation}
    \Lambda_{\pm1}^{\mu\nu}(\kk)=\frac{\lambda_\sigma}{6}e^{-ia_0k_y}\left[(1\mp3)\cos{\left(\tfrac{\sqrt{3}}{2}a_0k_x \right)} + \sqrt{3}(\gamma_{\mu\nu}\pm1)\sin{\left(\tfrac{\sqrt{3}}{2}a_0k_x \right)} + 2e^{ia_0k_y} \right],
\end{equation}
\end{subequations}
\end{widetext}
where we have defined $\gamma_{\uparrow \downarrow}=-\gamma_{\downarrow\uparrow}=1$. Moreover, $\Lambda_{n}^{\mu\mu}(\kk)=0$.

We now focus on momenta close to the KBZ $\Gamma$ point, and expand all expressions up to first order in $a_0\kk$ to obtain ($\hbar v_\sigma = \tfrac{3}{2}a_0t_0$)
\begin{subequations}
\begin{equation}
    \Phi_0(\kk) \approx -3t_0,
\end{equation}
\begin{equation}
    \Phi_{\pm1}(\kk)\approx \hbar v_\sigma (\mp k_x + i k_y),
\end{equation}
\begin{equation}
    \Lambda_0^{\uparrow\downarrow}(\kk) = -\left(\Lambda_0^{\downarrow\uparrow}(\kk) \right)^* \approx -\tfrac{\lambda_\sigma}{2}a_0(k_x-ik_y),
\end{equation}
\begin{equation}
    \Lambda_1^{\uparrow\downarrow}(\kk) = - \left( \Lambda_{-1}^{\downarrow\uparrow}(\kk) \right)^*\approx \tfrac{\lambda_\sigma}{2}a_0(k_x+ik_y),
\end{equation}
\begin{equation}
    \Lambda_{-1}^{\uparrow\downarrow}(\kk) = \Lambda_{1}^{\downarrow\uparrow}(\kk) \approx \lambda_\sigma.
\end{equation}
\end{subequations}
For clarity, we now use the basis ordering chosen in Ref.\ \cite{Gamayun_2018} for each spin quantum number, $\psi_{\kk}'=(\psi_{\kk,\uparrow}',\,\psi_{\kk,\downarrow})^T$,  with ($s=\uparrow,\downarrow$)
\begin{equation*}
    \psi_{\kk,s}=(-b_{\kk-\GG,s},a_{\kk-\GG,s},a_{\kk+\GG,s},b_{\kk+\GG,s})^T,
\end{equation*}
such that the effective Hamiltonian adopts the form
\begin{equation}
    \Hcal_{pq}''(\kk) = \begin{pmatrix}
        \Hcal_{0}^{pq}(\kk) & 0_{4\times4}\\
        0_{4\times4} & \Hcal_{0}^{pq}(\kk)
    \end{pmatrix} + \begin{pmatrix}
        0_{4\times4} & \Hcal_R^{pq}(\kk) \\
        \Hcal_R^{pq\dagger}(\kk) & 0_{4\times4}
    \end{pmatrix}
\end{equation}
where
\begin{equation}
    \Hcal_0^{pq}(\kk) = \begin{pmatrix}
        \hbar v_\sigma \kk\cdot\ssigma & \Delta Q_n(\kk) \\
        \Delta^* Q_n^\dagger(\kk) & \hbar v_\sigma \kk\cdot\ssigma
    \end{pmatrix}
\end{equation}
is the usual Kekul\'e graphene Hamiltonian for either a Kek-O ($Q_0=3t_0\sigma_z$) or a Kek-Y [$Q_{\pm1}=\hbar v_\sigma (\pm k_x-ik_y)\sigma_0$] texture. The Kekul\'e-RSO term $\Hcal_R^{pq}(\kk)$ has the general form
\begin{equation*}
    \Hcal_R^{pq}(\kk)=\begin{pmatrix}
        0 & i (\Delta_{-1}^{\downarrow\uparrow}(\kk))^* & i \xi(\Delta_{-n}^{\downarrow\uparrow}(\kk))^* & 0\\
        -i \Delta_{-1}^{\uparrow\downarrow}(\kk) & 0 & 0 & i \xi \Delta_{n}^{\uparrow\downarrow}(\kk) \\
        -i \xi^* \Delta_{-n}^{\uparrow\downarrow}(\kk) & 0 & 0 & i \Delta_{1}^{\uparrow\downarrow}(\kk) \\
        0 & -i \xi^* (\Delta_{n}^{\downarrow\uparrow}(\kk))^* & -i (\Delta_{1}^{\downarrow\uparrow}(\kk))^* & 0
    \end{pmatrix},
\end{equation*}
and simplifies as follows for the three possible values $n=-1,0,1$: For $n=0$ we get the Kek-O effective model
\begin{widetext}
\begin{equation}\label{eq:kekoR}
    \Hcal_R^{n=0}(\kk)=\begin{pmatrix}
        0 & -i\tfrac{\lambda_\sigma}{2} a_0 (k_x+ik_y) & i\xi \tfrac{\lambda_\sigma}{2} a_0 (k_x-ik_y) & 0\\
        -i\lambda_\sigma & 0 & 0 & -i\xi \tfrac{\lambda_\sigma}{2} a_0 (k_x-ik_y)\\
        i\xi^* \tfrac{\lambda_\sigma}{2} a_0 (k_x-ik_y) & 0 & 0 & i\tfrac{\lambda_\sigma}{2} a_0 (k_x+ik_y)\\
        0 & -i\xi^* \tfrac{\lambda_\sigma}{2} a_0 (k_x-ik_y) & -i\lambda_\sigma & 0
    \end{pmatrix},
\end{equation}
whereas for $n=\pm1$ we obtain the Kek-Y effective models
\begin{subequations}\label{eq:kekyR}
\begin{equation}
    \Hcal_R^{n=1}(\kk)=\begin{pmatrix}
        0 & -i\tfrac{\lambda_\sigma}{2} a_0 (k_x+ik_y) & -i\xi\tfrac{\lambda_\sigma}{2}a_0(k_x+ik_y) & 0\\
        -i\lambda_\sigma & 0 & 0 & i\xi\tfrac{\lambda_\sigma}{2}a_0(k_x+ik_y) \\
        -i\xi^* \lambda_\sigma & 0 & 0 & i\tfrac{\lambda_\sigma}{2} a_0 (k_x+ik_y)\\
        0 & -i\xi^* \lambda_\sigma  & -i\lambda_\sigma & 0
    \end{pmatrix},
\end{equation}
\begin{equation}
    \Hcal_R^{n=-1}(\kk)=\begin{pmatrix}
        0 & -i\tfrac{\lambda_\sigma}{2} a_0 (k_x+ik_y) & i\xi\lambda_\sigma & 0\\
        -i\lambda_\sigma & 0 & 0 & i\xi\lambda_\sigma \\
        -i\xi^* \tfrac{\lambda_\sigma}{2} a_0 (k_x+ik_y) & 0 & 0 & i\tfrac{\lambda_\sigma}{2} a_0 (k_x+ik_y)\\
        0 & i\xi^* \tfrac{\lambda_\sigma}{2} a_0 (k_x+ik_y) & -i\lambda_\sigma & 0
    \end{pmatrix}.
\end{equation}
\end{subequations}
\end{widetext}
Equations \eqref{eq:keky} and \eqref{eq:keko} are obtained from Eqs.\ \eqref{eq:kekyR} and \eqref{eq:kekoR}, respectively, by reordering the basis according to Eq.\ \eqref{eq:thisbasis}, and neglecting the terms linear in momentum.

\section{Landau levels Hamiltonian for the Kek-Y case}\label{app:LLHamiltonian}
For the Kekul\'e-Y case, the electronic Hamiltonian is obtained from \eqref{eq:keky} by substituting\cite{LuttingerKohn,LandauQM}
\begin{equation*}
\begin{split}
    \hbar v_\sigma (k_x + i k_y) \longrightarrow&\, v_\sigma(B) a^\dagger,\\
    \hbar v_\sigma (k_x - i k_y) \longrightarrow&\, v_\sigma(B) a,
\end{split}
\end{equation*}
with $v_\sigma(B) \equiv v_\sigma \sqrt{\frac{2e\hbar B}{c}}$ (in Gaussian units), and $a$ and $a^\dagger$ the LL ladder operators. Ordering the basis as
\begin{equation*}
    \begin{split}
        \{& |\uparrow,1,-1  \rangle,\,|\uparrow,-1,1 \rangle,\,|\downarrow,1,-1 \rangle,\,|\downarrow,-1,1 \rangle,\\
        &|\uparrow,-1,-1 \rangle,\,|\uparrow,1,1 \rangle,\,|\downarrow,-1,-1 \rangle,\,|\downarrow,1,1 \rangle \}
    \end{split}
\end{equation*}
where the quantum numbers correspond to spin ($s$), pseudo-spin ($\sigma$) and valley ($\tau$), respectively, the Landau level Hamiltonian takes the form
\begin{equation}
    H_{\mathcal{KY}}^{LL}=\begin{pmatrix}
    0 & h(B) \\
    h^\dagger(B) & 0
    \end{pmatrix},
\end{equation}
with
\begin{equation}
    h(B) = \begin{pmatrix}
    v_\sigma(B) a & \Delta v_\sigma(B) a^\dagger & 0 & -i\xi \lambda_\sigma\\
    \Delta v_\sigma(B) a & v_\sigma(B) a^\dagger & 0 & -i \lambda_\sigma \\
    i\lambda_\sigma & 0 & v_\sigma(B) a & \Delta v_\sigma(B) a^\dagger \\
    i\xi \lambda_\sigma & 0 & \Delta v_\sigma(B) a & v_\sigma(B) a^\dagger
    \end{pmatrix}.
\end{equation}
Defining the Landau level eigenstates $|s,\sigma,\tau; \ell \rangle$ ($\ell=0,1,2,\ldots$), we may propose a general solution of the form
\begin{equation}
    | \psi_\ell \rangle = \begin{pmatrix}
    |\ell-2 \rangle \\
    |\ell-2 \rangle \\
    |\ell-1 \rangle \\
    |\ell-1 \rangle \\
    |\ell-1 \rangle \\
    |\ell-3 \rangle \\
    |\ell \rangle \\
    |\ell-2 \rangle
    \end{pmatrix},
\end{equation}
such that for $\ell \ge 3$ we get
\begin{equation}\label{eq:LLnumerical}
    H_{\mathcal{KY}}^{LL}| \psi_\ell \rangle = \begin{pmatrix} 0_{4\times 4} & h_n(B) \\ h_n^\dagger (B) & 0_{4\times 4} \end{pmatrix}| \psi_\ell \rangle,
\end{equation}
with
\begin{widetext}
\begin{equation}
    h_\ell(B) = \begin{pmatrix}
    v_\sigma(B)\sqrt{\ell-1} & \Delta v_\sigma(B) \sqrt{\ell-2} & 0 & -i\xi \lambda_\sigma\\
    \Delta v_\sigma(B) \sqrt{\ell-1} & v_\sigma(B)\sqrt{\ell-2} & 0 & -i\lambda_\sigma \\
    i\lambda_\sigma & 0 & v_\sigma(B)\sqrt{\ell} & \Delta v_\sigma(B) \sqrt{\ell-1}\\
    i\xi\lambda_\sigma & 0 & \Delta v_\sigma(B)\sqrt{\ell} & v_\sigma(B) \sqrt{\ell-1}
    \end{pmatrix},\quad \ell\ge3.
\end{equation}
\end{widetext}
Setting $\ell=2$ yields the reduced problem
\begin{equation}\label{eq:Hforneq2}
    H_{\mathcal{KY}}^{LL}| \psi_2\rangle = \begin{pmatrix}
    0_{4\times 4} & h_2(B)\\
    h_2^\dagger (B) & 0_{3\times 3}
    \end{pmatrix} |\psi_2 \rangle,
\end{equation}
with
\begin{equation}
    h_2(B) = \begin{pmatrix}
    v_\sigma(B) & 0 & -i\xi \lambda_\sigma \\
    \Delta v_\sigma(B) & 0 & -i\lambda_\sigma\\
    i\lambda & \sqrt{2}v_\sigma(B) & \Delta v_\sigma(B) \\
    i\xi \lambda_\sigma & \sqrt{2}\Delta v_\sigma(B) & v_\sigma(B)
    \end{pmatrix}
\end{equation}
\begin{widetext}
and the reduced basis
\begin{equation}
      \{ |\uparrow,1,-1;0  \rangle,\,|\uparrow,-1,1;0 \rangle,\,|\downarrow,1,-1;1 \rangle,\,|\downarrow,-1,1;1 \rangle,\,|\uparrow,-1,-1;1 \rangle,\,|\downarrow,-1,-1;2 \rangle,\,|\downarrow,1,1;0 \rangle \}
\end{equation}
From this case we can extract the valley-sublattice-locked ($\tau \sigma = -1$) zero-energy mode
\begin{equation}
    | z_- \rangle = N_z^{-\tfrac{1}{2}}\Big[ -(\Delta \alpha - \beta)|\uparrow,1,-1;0 \rangle + (\alpha - \xi \beta)|\uparrow,-1,1;0 \rangle+ i \Delta \gamma|\downarrow,1,-1;1 \rangle - i\gamma|\downarrow,-1,1;1 \rangle \Big],
\end{equation}
\end{widetext}
where
\begin{subequations}
\begin{equation}
    \alpha = v_\sigma^2(B)(1-\Delta^2),
\end{equation}
\begin{equation}
    \beta = \lambda_\sigma^2(\xi - \Delta),
\end{equation}
\begin{equation}
    \gamma = \lambda_\sigma v_\sigma(B)\Delta(1-\xi\Delta),
\end{equation}
\begin{equation}
    N_z =(\alpha^2+\gamma^2)(1+\Delta^2) + \beta^2(1+\xi^2) - 2(\xi+\Delta)\alpha\beta .
\end{equation}
\end{subequations}

Then, setting $\ell=1$ we obtain the reduced problem
\begin{equation}\label{eq:ZeroModel}
    H_{\mathcal{KY}}^{LL}|\psi_{1} \rangle = \begin{pmatrix} 0_{2\times2} & h_1(B) \\ h_1^\dagger(B) & 0_{2\times 2}  \end{pmatrix}|\psi_{1} \rangle,
\end{equation}
with
\begin{equation}
    h_1(B)=\begin{pmatrix}
    i\lambda_\sigma & v_\sigma(B) \\
    i\xi \lambda_\sigma & \Delta v_\sigma(B)
    \end{pmatrix},
\end{equation}
and the reduced basis
\begin{equation*}
\{|\downarrow,1,-1;0 \rangle, |\downarrow,-1,1;0 \rangle,|\uparrow,-1,-1;0 \rangle, |\downarrow,-1,-1;1 \rangle \}.    
\end{equation*}
This can be diagonalized analytically, and gives the eigenvalues ($\eta,\zeta = \pm 1$)
\begin{widetext}
\begin{equation}\label{eq:ZeroModes}
    \varepsilon_{\eta,\zeta}(B) = \frac{\eta}{\sqrt{2}}  \sqrt{(1+\Delta^2)v_\sigma^2(B) + (1+\xi^2)\lambda_\sigma^2+\zeta\sqrt{\left[(1+\Delta^2)v_\sigma^2(B)+(1+\xi^2)\lambda_\sigma^2 \right]^2 - 4\lambda_\sigma^2 v_\sigma^2(B) (\Delta - \xi)^2}}.
\end{equation}
\end{widetext}
Two additional zero modes are recovered in the case of $\xi = \Delta$, corresponding to the eigenstates
\begin{equation*}
\begin{split}
    | z_+' \rangle =& \frac{v_\sigma(B) | \uparrow,-1,-1;1 \rangle - i\lambda_\sigma | \downarrow,-1,-1;0 \rangle}{\sqrt{v_\sigma^2(B)+\lambda_\sigma^2}},\\
    | z_-' \rangle =& \frac{-\Delta |\downarrow,1,-1;0 \rangle + | \downarrow,-1,1;1 \rangle}{\sqrt{1+\Delta^2}},
\end{split}
\end{equation*}
which split as
\begin{equation}\label{eq:zeromodesplitting}
    \delta\varepsilon_{z'} = \frac{2\lambda_\sigma v_\sigma(B)}{\sqrt{(1+\Delta^2)v_\sigma^2(B) + (1+\xi^2)\lambda_\sigma^2}} |\Delta - \xi |
\end{equation}
for $0<|\Delta - \xi|\ll1$.

\begin{figure}[t!]
    \centering
    \includegraphics{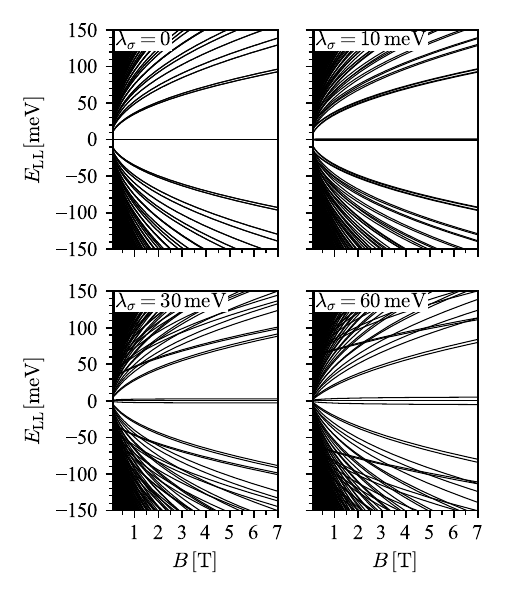}
    \caption{Landau levels for different values of $\lambda_\sigma$, keeping $\Delta=0.1$ fixed. Note the splitting of the zero-energy quadruplet for $\lambda_\sigma\ne0$, into a zero-energy doublet and two dispersive LLs.}
    \label{fig:LLKekRashba}
\end{figure}

Finally, setting $\ell=0$ yields the zero-energy mode
\begin{equation}
    | z_+ \rangle = |\downarrow,-1,-1;0 \rangle.
\end{equation}

Note that in the absence of Rashba spin-orbit coupling $(\lambda_\sigma\rightarrow 0)$, the four zero-energy modes identified reduce to those reported by Gamayun \emph{et al.} in Ref.\ \onlinecite{Gamayun_2018}:
\begin{equation*}
\begin{split}
    \left. | z_+  \rangle\right|_{\lambda_\sigma=0} =& | \downarrow,-1,-1;0 \rangle,\\
    \left. | z_+' \rangle\right|_{\lambda_\sigma=0} =& | \uparrow,-1,-1;1 \rangle,\\
    \left. | z_- \rangle\right|_{\lambda_\sigma=0} =& \frac{-\Delta | \uparrow,1,-1;0 \rangle + |\uparrow,-1,1;0 \rangle}{\sqrt{1+\Delta^2}},\\
    \left. | z_-' \rangle\right|_{\lambda_\sigma=0} =& \frac{-\Delta | \downarrow,1,-1;0 \rangle + |\downarrow,-1,1;1 \rangle}{\sqrt{1+\Delta^2}}.
\end{split}
\end{equation*}

\section{Landau level spectrum of Kek-Y graphene with RSO coupling}\label{app:LLsKekYandRSO}
Figure \ref{fig:LLKekRashba} shows the LL spectra of Kek-Y graphene, with fixed Kekul\'e parameter $\Delta=0.1$, and for several values of the RSO term $\lambda_\sigma$, keeping the RSO distortion $\xi=0$. At first sight, this LL spectrum resembles that of RSO graphene (Fig.\ \ref{fig:LLJustRashba}), with a duplicated fan due to the two Fermi velocities introduced by the Kek-Y distortion (Fig.\ \ref{fig:LLJustKek}). However, a major qualitative difference with those cases is that the zero-energy quadruplet breaks breaks in the case of simultaneous Kek-Y and RSO effects, into a zero-energy doublet surrounded by two dispersive LLs that split according to Eq.\ \eqref{eq:zeromodesplitting}.

\section{Magnetic field splittings in the Fermi-level dependent DOS of pure Kek-Y and RSO samples}\label{app:asymptotic}
The high-DOS fans in Fig.\ \ref{fig:splittings} are direct visualizations of constant energy cuts of the LL fans of Figs.\ \ref{fig:LLJustRashba} and \ref{fig:LLJustKek}, for $E_{\rm LL} = \varepsilon_F$. The $\varepsilon_F$-dependent splittings can be extracted directly from the $B$-dependent eigenvalues of the LL Hamiltonians \eqref{eq:LLnumerical}, \eqref{eq:Hforneq2} and \eqref{eq:ZeroModel}. For simplicity, let us focus on the splitting between the rightmost splitting, indicated in Fig.\ \ref{fig:splittings}(h) with the symbol $\blacksquare$.

In the case of only a Kek-Y distorsion ($\Delta\ne0$, $\lambda_\sigma=0$, $\xi=0$), this splitting occurs between the LLs
\begin{equation}
\begin{split}
    \varepsilon_4^{(1)}(B)=&v_\sigma(B) \sqrt{1+\Delta^2},\\
    \varepsilon_5^{(2)}(B)=&v_\sigma(B)\sqrt{\frac{3(1+\Delta^2)-\sqrt{1+34\Delta^2+\Delta^4}}{2}},
\end{split}
\end{equation}
where $\varepsilon_m^{(\ell)}$ is the $m$th eigenvalue, by increasing energy, obtained from the model \eqref{eq:LLnumerical} by fixing the LL index $\ell$. The magnetic field splitting between these two LLs at fixed $\varepsilon_F$ is obtained as $\delta B = B_> - B_<$, where
\begin{equation}\label{eq:FixedEF}
    \varepsilon_5^{(2)}(B_>) = \varepsilon_4^{(1)}(B_<) = \varepsilon_F.
\end{equation}
This gives
\begin{equation}
\begin{split}
    \delta B =& \frac{\varepsilon_F^2}{\hbar^2v_\sigma^2}\sqrt{\frac{c}{2e\hbar}}\\
    &\times\left[\frac{2}{3(1+\Delta^2)-\sqrt{1+34\Delta^2+\Delta^4}} - \frac{1}{1+\Delta^2} \right]\\
    = & \frac{8\Delta^2}{\hbar^2v_\sigma^2}\varepsilon_F^2\sqrt{\frac{c}{2e\hbar}} + \mathcal{O}\{\Delta^4\},
\end{split}
\end{equation}
leading to Eqs.\ \eqref{eq:powerlaw} and \eqref{eq:Acoeff}.

We may follow the same procedure in the case of only RSO coupling ($\Delta=0$, $\lambda_\sigma\ne0$,\,$\xi=0$), where the relevant LL energies are
\begin{equation}
\begin{split}
    \varepsilon_4^{(1)}(B)=&\sqrt{\hbar^2v_\sigma^2(B) + \lambda_\sigma^2},\\
    \varepsilon_5^{(2)}(B)=&\sqrt{\frac{3v_\sigma^2(B) + \lambda_\sigma^2 - \sqrt{v_\sigma^4(B) + 6\lambda_\sigma^2v_\sigma^2(B) + \lambda_\sigma^4}}{2}}.
\end{split}
\end{equation}
In this case, Eq.\ \eqref{eq:FixedEF} gives $B_< = \tfrac{\varepsilon_F^2-\lambda_\sigma^2}{\hbar^2v_\sigma^2}\sqrt{\tfrac{c}{2e\hbar}}$, whereas for $B_>$ we get the equation
\begin{equation*}
\begin{split}
    2\varepsilon_F^2 =& \lambda_\sigma^2 + (\hbar v_\sigma)^2\sqrt{\frac{c}{2e\hbar}}B\\
    &\times\left[3 - \sqrt{1 + \frac{6\lambda_\sigma^2}{(\hbar v_\sigma)^2B}\sqrt{\frac{c}{2e\hbar}} + \frac{\lambda_\sigma^4}{(\hbar v_\sigma)^4 B^2}\frac{c}{2e\hbar}}  \right].
\end{split}
\end{equation*}
This can be simplified by working in the limit of $\tfrac{\lambda_\sigma^2}{v_\sigma^2(B)}\ll1$, and expanding up to second order. This gives the quadratic equation
\begin{equation*}
    B^2 - \frac{\varepsilon_F^2 + \lambda_\sigma^2}{\hbar^2v_\sigma^2}\sqrt{\frac{c}{2e\hbar}}B + \frac{2\lambda_\sigma^4}{\hbar^2v_\sigma^4}\frac{c}{2e\hbar} \approx 0,
\end{equation*}
from which we take the solution
\begin{equation*}
    B_> \approx \frac{\varepsilon_F^2+\lambda_\sigma^2}{2\hbar^2v_\sigma^2}\left[1+\sqrt{1-\frac{8\lambda_\sigma^4}{(\varepsilon_F^2+\lambda_\sigma^2)^2}} \right].
\end{equation*}
In the limit of $\tfrac{\lambda_\sigma}{\varepsilon_F}\ll1$ we may approximate $B_> \approx \tfrac{\varepsilon_F^2+\lambda_\sigma^2}{2\hbar^2v_\sigma^2}$, yielding the magnetic field splitting
\begin{equation}
    \delta B = \frac{2\lambda_\sigma^2}{\hbar^2v_\sigma^2}\sqrt{\frac{c}{2e\hbar}} + \mathcal{O}\{ (\tfrac{\lambda_\sigma}{\varepsilon_F})^2\},
\end{equation}
leading to Eq.\ \eqref{eq:saturation}.

\bibliographystyle{unsrt}
\bibliography{biblio}

\end{document}